\documentclass[12pt]{article}
\usepackage{amsmath,amssymb}

\newtheorem{prop}{Proposition}

\begin{document}

\begin{center}

\section*{Small Oscillations of a Vortex Ring: Hamiltonian Formalism and Quantization}

{S.V. TALALOV}

{Department of Applied Mathematics, Togliatti State University, \\ 14 Belorusskaya str.,
 Tolyatti, Samara region, 445020 Russia.\\
svt\_19@mail.ru}

\end{center}

\begin{abstract}
This article investigates small oscillations of a vortex ring with zero thickness that evolves under the Local Induction Equation (LIE). We deduce the differential equation that describes the dynamics of these oscillations.  We suggest the new approach to the Hamiltonian description of this dynamic system. This approach is based on the extension of the set of dynamical variables by adding    the 	circulation	$\Gamma$ as a dynamical variable. 		The constructed theory is invariant under the transformations of the Galilei group.  The appearance of this group allows for a  new viewpoint on the energy of a  vortex filament with zero thickness. We quantize this dynamical system and calculate the spectrum of the energy and acceptable circulation values. 
	The physical states of the theory are constructed with help of coherent states for the Heisenberg -Weyl group.
\end{abstract}

{\bf keywords:}   vortex ring, constrained Hamiltonian systems,  quantization

 {\bf PACS numbers:}   47.10.Df    47.32.C

\vspace{5mm}

\section{Introduction}	

 The study of  various vortex structures has a long history.  
In spite of this fact, 
  dynamics of  such objects  continue to attract interest\cite{Moff}.
	Without trying to make a review of the literature on this topic, we will mention only some works that are directly relevant to our research.
	So,  in 	 the work \cite{Hasim}  the vortex filament (in the LIE approximation)
		was described firstly  in terms of solutions of non-linear Schr\"odinger equation.  It should also be mentioned that  		
		   gauge equivalence     between the non-linear Schr\"odinger equation and the continuous Heisenberg    spin chain exists\cite{TakFad}.
It is well-known that quantization of  similar  non-linear systems is a complicated problem  (see, for example, \cite{Sklyanin}). As a consequence, the connection with the initial hydrodynamical system (the vortex ring in our case)  maybe  seems not obvious.  That is why the investigations of the small  perturbations of certain initial and stable configurations are interesting, both classical and quantum. 
Let us note here the work \cite{Majda_Bertozzi}, where, in particular, the system of  small-perturbed  straightforward vortex filaments with  certain interactions  were investigated. 
In the work \cite{AbhGuh}, the quantization of such filaments was considered.
It is also necessary to mention the direction of the study of quantum vortices in superfluid helium \cite{Donn,Aarts,Andr}. Since our research lies in a different plane, we will not dwell on this in detail.

 In this work, we  consider the  closed evolving curve ${\boldsymbol{r}}(\tau,\xi)$  that is defined by the formula
\begin{equation}
        \label{involve}
                                   {\boldsymbol{r}}(\tau,\xi) =  \boldsymbol{q} + 
         {R_0}\, \int\limits_{0}^{2\pi}  \left[\, {\xi - \eta}\,\right] {\boldsymbol j}(\tau,\eta) d\eta\,.  
                  \end{equation}

Both parameter $\xi$  that  parametrizes the curve  ${\boldsymbol{r}}(\cdot,\xi)$ and evolution parameter  $\tau$ 
 are dimensionless parameters.     The constant $R_0$   defines the scale of length.
								The notation $[\,x\,]$ means the integer part of the number $x/{2\pi}$, the  variables  $\boldsymbol{q} = \boldsymbol{q}(\tau)$  may be $\tau$-depended
								(conditionally, these are the coordinates   of the ''mass center''). 											
								$2\pi$-periodical   vector function ${\boldsymbol j}(\xi) \in E_3$ defines unit tangent vector.  We postulate that 
								function  ${\boldsymbol{r}}(\tau,\xi)$ satisfies the LIE equation
								\begin{equation}
        \label{LIE_eq}
        \partial_\tau {\boldsymbol{r}}(\tau ,\xi) = \frac{1}{R_0}\,
        \partial_\xi{\boldsymbol{r}}(\tau ,\xi)\times\partial_\xi^{\,2}{\boldsymbol{r}}(\tau ,\xi)\,.
        \end{equation}
								
					Consequently, the  								function  ${\boldsymbol j}(\xi) $   satisfies the equation for the 
								continuous Heisenberg spin chain

\begin{equation}
        \label{CHSCeq}
        \partial_\tau {\boldsymbol{j}}(\tau ,\xi) = 
        \,{\boldsymbol{j}}(\tau ,\xi)\times\partial_\xi^{\,2}{\boldsymbol{j}}(\tau,\xi)\,.
        \end{equation}

               The following   equalities are fulfilled too:
            \begin{equation}
  \label{constr_j_0}
    \int\limits_{0}^{2\pi}{j}_k(\xi)\,d\xi  = 0\, \qquad  (k=x,y,z)\,,
   \end{equation}

			The original      space-time symmetry group  for this system is the group  $E(3)\times  E_\tau$,
 where the  group  $E(3)$ is the group of motions of space $E_3$ and 
			$E_\tau$ is the group of ''translations'' $\tau \to \tau +c$.

The   standard simplest configuration here will be the vortex ring with radius $R_0$ that moves parallel to $z$-axis with   some constant velocity:
\begin{equation}
        \label{ring_1}
  {\boldsymbol{r}_0}(\tau,\xi) = 	\boldsymbol{q}_0	+ 	{R_0}\, \int\limits_{0}^{2\pi}  \left[\, {\xi - \eta}\,\right] {\boldsymbol j}_0(\tau,\eta) d\eta\,.													
		\end{equation}
   Tangent       vector  ${\boldsymbol j}_0$  has the following coordinates:
\begin{equation}
        \label{tang_v}
                                   {\boldsymbol{j}_0}(\xi) = \{ - \sin\xi\,, \quad \cos\xi\,, \quad 0\}\,.
\end{equation}
The quantities $ {q_0}_i = const$ for the indexes  $i = x,y$ here;  
the  quantity ${q_0}_z$ is some linear function on the variable $\tau$ that will be specified latter. 
	
		We consider the small perturbation $(\varepsilon << 1)$   of the tangent vector (\ref{tang_v})  and coordinates $  \boldsymbol{q}_0$:
		\begin{equation}
        \label{tang_v_per}
          \boldsymbol{q}  =  \boldsymbol{q}_0 + \varepsilon\, \boldsymbol{q}_{prt}\,,\qquad 																
																\boldsymbol{j}(\tau,\xi)  = {\boldsymbol{j}_0}(\xi) + \varepsilon {\boldsymbol{j}_{prt}}(\tau,\xi) \,. 
\end{equation}
		Therefore, we have representation 
		$${\boldsymbol{r}}(\tau ,\xi) = {\boldsymbol{r}_0}(\tau,\xi)   + \varepsilon  {\boldsymbol{r}_{prt}}(\tau ,\xi)\,.$$

		Let us substitute the representation (\ref{tang_v_per}) for the vector  ${\boldsymbol{j}}(\xi)$   into equation  (\ref{CHSCeq}). Taking into account the equality 
		$\partial^{\,2}_\xi {\boldsymbol{j}_0}(\xi) = - {\boldsymbol{j}_0}(\xi)$ and neglecting the terms of order $\varepsilon^2$,   we deduce the equation
		
		\begin{equation}
        \label{lin_eq}
        \partial_\tau {\boldsymbol{j}_{prt}}(\tau ,\xi) =  {\boldsymbol{j}_0}(\tau ,\xi) \times 
				\left[\, {\boldsymbol{j}_{prt}}(\tau ,\xi)  +   \partial_\xi^{\,2}{\boldsymbol{j}_{prt}}(\tau,\xi)\, \right]\,.
				        \end{equation}
		Thus, we have the linear equation that describes the dynamics of small perturbations for the vortex ring  (\ref{ring_1}).
		Next, we will not write the ${''prt''}$ index explicitly, hoping that this will not lead to misunderstandings.
				It must be emphasized that equalities (\ref{constr_j_0}) must be fulfilled strongly for all configurations (perturbed or not perturbed).

		The symmetry of the initial configuration ${\boldsymbol{j}_0}(\tau ,\xi)$ makes natural using the  cylindrical coordinates $\{\rho, \phi, z\}$.
		The z-axis of such a system coincides with the axis of the unperturbed vortex ring. 		
		Let three vectors $\{ {\boldsymbol{e}_\rho}\,,  {\boldsymbol{e}_\phi}\,, {\boldsymbol{e}_z}\,\}$ denote the local basis of the cylindrical system.
		Obviously, in  special case when $	\boldsymbol{j} \equiv {\boldsymbol{j}_0}$, the parameter $\xi = \phi$.  Therefore,
		$$ {\boldsymbol{j}_0} =  {\boldsymbol{e}_\phi}\,.$$
    Because we consider the small perturbations of  the function ${\boldsymbol{j}_0}$ only, we assume  $\xi = \phi$ for all configurations.

		Thus, the equation (\ref{lin_eq}) takes the following form in the cylindrical basis:
\begin{equation}
        \label{lin_eq_cyl}
    \partial_\tau {\boldsymbol{j}}(\tau ,\xi) =  \Bigl({{j}_z}(\tau ,\xi)  +   \partial_\xi^{\,2}{{j}_z}(\tau,\xi)\Bigr) {\boldsymbol{e}_\rho} 
- \Bigl(   \partial_\xi^{\,2}{{j}_\rho}(\tau,\xi)  -2\, \partial_\xi {j}_\phi (\tau,\xi)  \Bigr) \boldsymbol{e}_z  \,.
\end{equation}
This equation demonstrates  that $ \partial_\tau {j_\phi}(\tau ,\xi) \equiv 0\,.$
For example, the initial data ${j_\phi}(0    ,\xi) \equiv j_\phi^{\,0} $, where  $  \partial_\xi j_\phi^{\,0} \equiv 0$,  lead to the  equality 
$${j_\phi}(\tau ,\xi) \equiv j_\phi^{\,0}\,, \qquad  j_\phi^{\,0}  = const\,. $$
 Obviously, the perturbation   ${\boldsymbol{j}_0}(\xi) \to {\boldsymbol{j}_0}(\xi) + \varepsilon j_\phi^{\,0}{\boldsymbol{e}_\phi}$  
 conserves the form of   the original vortex    ${\boldsymbol{r}_0}$.   In our subsequent considerations we will consider the case $j_\phi^{\,0}  = const $  only.
 Therefore, the perturbation amplitude   ${\boldsymbol j}(\tau ,\xi)$   of the vector  ${\boldsymbol{j}_0}(\tau ,\xi)$ has following form in the cylindrical basis:
\begin{equation}
        \label{j-two}
{\boldsymbol j}(\tau ,\xi)   
  =   j_\rho(\tau,\xi){\boldsymbol{e}}_\rho  +    j_\phi^{\,0} \boldsymbol{e}_\phi +    j_z(\tau ,\xi)  {\boldsymbol{e}_z}\,.
\end{equation}

The following  identity takes place:
\begin{equation}
        \label{iden_1}
				{\boldsymbol j}(\tau ,\xi){\boldsymbol{j}_0}(\tau ,\xi) \equiv j_\phi^{\,0}\,.
	\end{equation}

The notation ${\boldsymbol j}{\boldsymbol{j}_0}$ means an inner product of two vectors.
In the future, it is more convenient for us to consider the non-trivial components of 
the vector ${\boldsymbol j}$ in a complex-valued  form. 
We  introduce the notation ${\sf j}$  for the complex perturbation amplitude to avoid any ambiguity.  Thus,
$$  {\sf j}  =  j_\rho  + i  j_z\,.$$
Equation (\ref{lin_eq_cyl}) writing in complex form as follows:
\begin{equation}
        \label{lin_eq_com}
\partial_\tau {\sf j} =  - i  \partial_\xi^{\,2} {\sf j}    - \frac{i}{2}\Bigl({\sf j}   - \overline{\,\sf j\,}\, \Bigr)   \,.
\end{equation}

This simple equation can be solved explicitly:

\begin{equation}
        \label{sol_1}
				{\sf j}(\tau ,\xi) = \sum_{n} {\sf j}_{\,n}\, e^{\,i\,[n\xi  +  n \sqrt{n^2 - 1}\,\tau\,]} \,,   
\end{equation}			

where ${\sf j}_{\,n} \equiv const$ and the coefficients $ \overline{\,	{\sf j}\,}_{\,-n}$  and  ${\sf j}_{\,n}$ are related to each other as follows:

\begin{equation}
        \label{jn_jn-conj}
			\overline{\,	{\sf j}\,}_{\,-n}     =    2 \left[n\sqrt{n^2 -1} - n^2 + \frac{1}{2} \right]  {\sf j}_{\,n} \,.
	\end{equation}

	It is clear  that  this solution is stable for $\tau \to \infty$.  			
			Note  that the stability of the real vortex rings is a non-trivial problem in general. For example, this problem was investigated in the work 
			\cite{Protas} for the Norbury rings.

							Restrictions  (\ref{constr_j_0}) are rewriting in terms of the cylindrical coordinates as follow:	
			\begin{equation}
        \label{cons_cyl_1}
				\int_0^{2\pi} j_\rho(\xi)\,e^{\pm i\xi}d\xi =0\,,\qquad   \int_0^{2\pi} j_z(\xi)\,d\xi =0\,.
				\end{equation}
			
		As regards to the coefficients 	${\sf j}_{\,0}$,  ${\sf j}_{\pm 1}$ in the solution (\ref{sol_1}),  constraints    (\ref{cons_cyl_1}) lead to the formulas
		\begin{eqnarray}
		\label{j_0}
		{\sf j}_{\,0} & = & \frac{1}{2\pi} \int_0^{2\pi}\Bigl(j_\rho  + i  j_z \Bigr)d\xi = \frac{1}{2\pi} \int_0^{2\pi}j_\rho \,d\xi\,, \\[2mm]
		{\sf j}_{\,\pm 1} & = & \frac{1}{2\pi}\int_0^{2\pi}\Bigl(j_\rho  + i  j_z \Bigr)\,e^{\mp i\xi} d\xi = \frac{1}{2\pi}\int_0^{2\pi}j_z (\pm\sin\xi + i\cos\xi)\,d\xi\,.
		\label{j_pm}
			\end{eqnarray}
		So, last equalities lead to the restrictions for the coefficients ${\sf j}_{\pm 1}$:
			\begin{equation}
        \label{j_pm_rstr}
			{\sf j}_{\, 1}  =  - \overline{\,{\sf j}\,}_{\,-1}\,.
			\end{equation}
					
			As regards the coefficient 	${\sf j}_{\,0}$,  the equalities (\ref{j_0}) lead to the restriction  ${\sf j}_{\,0}  =  \,\overline{\,{\sf j}\,}_{\,0}$.
			Although these restrictions are deduced from  the constraint (\ref{constr_j_0}), they are consistent with the formula (\ref{jn_jn-conj}).

				\section{Dynamical invariants and  extended set of the  variables} 
       
  The consideration of vortex structure in terms of the equation (\ref{LIE_eq}) only  is apparently too formal.
	 As an addition to the LIE equation, we postulate in our theory the canonical formulas for the momentum and  angular momenta, that was deduced in 	
	the fluid dynamics \cite{Batche}:
	  
   \begin{equation}
        \label{p_and_m_st}
        \tilde{\boldsymbol{p}} = \frac{1}{2 }\,\int\,\boldsymbol{r}\times\boldsymbol{\omega}(\boldsymbol{r})dV\,,\qquad 
         \tilde{\boldsymbol{s}} =\frac{1}{3}\, \int\,\boldsymbol{r}\times
         \bigl(\boldsymbol{r}\times\boldsymbol{\omega}(\boldsymbol{r})\bigr)dV\,.
        \end{equation}

The vector  $\boldsymbol{\omega}(\boldsymbol{r})$  means  the vorticity.  The fluid density   $\varrho \equiv 1$  here.

      As well-known,   the vorticity of the  closed vortex filament calculates by means of the formula
   \begin{equation}
        \label{vort_w}
     \boldsymbol{\omega}(\boldsymbol{r}) =  \Gamma
                  \int\limits_{0}^{2\pi}\,\hat\delta(\boldsymbol{r} - \boldsymbol{r}(\xi))\partial_\xi{\boldsymbol{r}}(\xi)d\xi\,,
       \end{equation}            
   where the symbol $\Gamma$ denotes the circulation  and the symbol  $\hat\delta(\xi)$ means $2\pi$-periodical $3D$  $\delta$-function.
	
	Taking into account the formulae  (\ref{involve}),   (\ref{constr_j_0}) and   (\ref{vort_w}),   the following  expression for the canonical momentum is deduced:
	\begin{equation}
        \label{impuls_def}
                 \tilde{\boldsymbol{p}}  =     {R_0}^2 \Gamma                  {\boldsymbol f} \,,
                 \qquad 
                  {\boldsymbol f} = \frac{1}{2}\iint\limits_{0}^{2\pi}  \left[\, {\xi - \eta}\,\right]\,{\boldsymbol j}(\eta)\times{\boldsymbol j}(\xi)d\xi  d\eta\,.
      \end{equation} 
	A similar formula can be written for the angular momenta $\tilde{\boldsymbol{s}}$;  however, we omit the relevant details here.
		As opposed to formulae for the values 	$\tilde{\boldsymbol{p}}$  and  $\tilde{\boldsymbol{s}}$, the canonical formula for the energy ${\mathcal E}$  gives the unsatisfactory result because of the divergence of the integral.  We will return to this question later.
						        
	Let  us substitute the expansion (\ref{tang_v_per})  into expression for the vector ${\boldsymbol f}$ (see (\ref{impuls_def})). Taking into account identity (\ref{iden_1}), 
	the following formula for the vector ${\boldsymbol f}$  holds:
	\begin{equation}
        \label{f_gen}
				{\boldsymbol f} = \pi(1 + 2\,\varepsilon j_\phi^{\,0}) {\boldsymbol e}_z  - \varepsilon  {\boldsymbol f}_{\!\!\perp} \,,\qquad    
								{\boldsymbol f}_{\!\!\perp} =  \int_0^{2\pi} j_z(\xi) {\boldsymbol e}_\phi \,d\xi\,.
				\end{equation}
	The quantity ${\boldsymbol f}$ includes both unperturbed  ($\pi{\boldsymbol e}_z$) and perturbed 
	($ 2\pi\,\varepsilon j_\phi^{\,0} {\boldsymbol e}_z  - \varepsilon  {\boldsymbol f}_{\!\!\perp}$) values.
		Correspondingly, the full  momentum $\tilde{\boldsymbol{p}}$  has following form:
	
	$$ \tilde{\boldsymbol{p}}   =  - \varepsilon \tilde{\boldsymbol{p}}_{\!\perp} + \tilde{p}_{\,\parallel} {\boldsymbol{e}_z}\,,$$
	where 
	\begin{equation}
        \label{p_z1}
	\tilde{p}_{\,\|} = \pi (1 + 2\,\varepsilon j_\phi^{\,0}) {R_0}^2 \Gamma \,.
	\end{equation}
	As for the amplitudes of perturbations for the momentum, we have the following formulas:

	\begin{eqnarray}
	\label{p_z_pert}
	 \tilde{p}_z & = & 2\pi   {R_0}^2 \Gamma j_\phi^{\,0} \, \\	     
				\tilde{\boldsymbol{p}}_{\!\perp} &  =  &  {R_0}^2 \Gamma  {\boldsymbol f}_{\!\!\perp}\,.
				\label{imp_perp}
				\end{eqnarray} 

	Next, we intend to construct a Hamiltonian dynamical system that corresponds  to the equation (\ref{lin_eq}).
	For the general case  (\ref{LIE_eq}),  the corresponding approach was proposed by the author in the work \cite{Tal_1}. We will briefly recall here the main points of the proposed theory, making the necessary modifications along the way.
        
				In our opinion, the following steps must be done to construct the physically interpreted dynamical system in our case:
				
								\begin{enumerate}
								      \item  As mentioned above, we need to supplement the equation (\ref{LIE_eq}) with 					
																						   formulas for the momentum $\tilde{\boldsymbol{p}}$  and the angular momenta $\tilde{\boldsymbol{s}}$;
         			 \item    The variable ''velocity of liquid'' is absent in our theory.	We propose to take into account the dynamics of the surrounding fluid in a minimal way:
			to declare the value $\Gamma$ as a dynamic variable, in addition to   variables $\boldsymbol{j}(\xi)$ and  $\boldsymbol{q}$.
					We  denote as ${\mathcal A}$ this (extended) set of the dynamical variables  
 $\{\,\boldsymbol{q}\,,{\boldsymbol j}(\xi)\,,\Gamma\,\}$ constrained by the conditions  (\ref{constr_j_0}).  
        \item The theory must contain a sufficient number of dimensional constants.
				So far, we use the dimensionless ''time'' $\tau$.   Consequently,   the additional dimensional  constant $t_0$ that defines the  scale of time,
				must be added to the theory in some way. 	Subsequently, we will express this constant in terms of other  dimensional constants that have a clear physical meaning in our model.
				In addition to the $R_0$ and $t_0$ constants, the theory should contain a ''mass constant'' $m_0$.  This constant will appear in the theory in a completely natural way later on.
				\end{enumerate}

	As a subtotal, we have the following
  
					\begin{prop}
										The set ${\mathcal A}$  parametrizes the considered dynamical system - the closed vortex filament
					${\boldsymbol{r}}(\xi) $ 	that  evolving in accordance with the LIE equation.   	This dynamical system has a momentum $\tilde{\boldsymbol{p}}$  and  angular momenta $\tilde{\boldsymbol{s}}$  that  calculated as prescribed above.
					\end{prop} 
	
	To perform  the hamiltonization of our system, we are going to describe the set ${\mathcal A}$ in terms of   other  variables.
 The reasons are following:
\begin{itemize}
\item we intend to expand the symmetry group  $E(3)\times  E_\tau$   to Galilei group  ${\mathcal G}_3$ and use the group-theoretical approach for the  definition of the energy of our system;
 \item new variables  will be more suitable for subsequent quantization that will be fulfilled later in this article. 
\end{itemize}
	
	In addition, the new variables will give the obvious interpretation of the considered dynamical system as some structured particle.
	It is probably appropriate to mention here the  pioneering work \cite{Thom} in which the observed particles are modeled by vortex structures.

	As a first step, we extend the set ${\mathcal A}$. Let us denote as  ${\mathcal A}^{\,\prime}$ the set of the independent variables
	$({\boldsymbol q}\,;\, \tilde{\boldsymbol p}\,; {\boldsymbol j}(\xi)\,)$. The formula  (\ref{impuls_def}) makes the injection $F$:
	
	$$F:\quad {\mathcal A} \quad \mathrel{\mathop{\longrightarrow}^{F} } \quad {\mathcal A}^{\,\prime}\,, \qquad  {\rm Ran}\, F \subset {\mathcal A}^{\,\prime}\,.  $$
	
	On the set ${\mathcal A}^{\,\prime}$ the action of the central extended  Galilei group $\widetilde{\mathcal G}_3 $ is defined by natural way.
	We parametrize the elements $g \in \widetilde{\mathcal G}_3 $ as follows:
	$$ g:\qquad \Bigl( {\mathcal R}\,, {\boldsymbol v}\,,  {\boldsymbol a}\,, c\,; m_0   \Bigr)\,$$
	where ${\mathcal R}\in SO(3)$,  ${\boldsymbol v}\,,  {\boldsymbol a} \in E_3$, $c \in {\sf R}$ and the central charge $m_0 \in {\sf R}$.  Traditionally, the last parameter is interpreted as  
	''mass of the particle''. Before determining the action of this group on the set ${\mathcal A}^{\,\prime}$,
	we introduce the factor $m_0/R_0^3$ in standard hydrodynamical formulas (\ref{p_and_m_st})  to provide   the dimension for the values $\tilde{\boldsymbol{p}}$ 
			and   $\tilde{\boldsymbol{s}}$ as  in classical mechanics:
			
			\begin{equation}
			\label{rep_factor}
			\tilde{\boldsymbol{p}} \longrightarrow  {\boldsymbol{p}} =  (m_0/R_0^3) \tilde {\boldsymbol{p}}\,,\qquad 
			\tilde{\boldsymbol{s}} \longrightarrow  {\boldsymbol{s}} =  (m_0/R_0^3) \tilde {\boldsymbol{s}}\,.
			\end{equation}
			
			Taking into account this redefinition, the group action $\circ$ on the set ${\mathcal A}^{\,\prime}$ is defined as follows:
	
	$$ g\circ ({\boldsymbol q}  \,;\, {\boldsymbol p}\,; {\boldsymbol j}(\xi)\,) =
	({\mathcal R}{\boldsymbol q} +  {\boldsymbol v}t_0\tau  + {\boldsymbol a}\,;\, {\mathcal R}{\boldsymbol p}  + m_0 {\boldsymbol v}\,; {\mathcal R}{\boldsymbol j}(\xi)\,)\,,$$
	and $g\circ (t_0\tau) = t_0\tau + c$. 
	
		Next, we  introduce the variables
  $$  q_i(0) =   q_i  - {\tau}(t_0/ m_0) p_i \,,\qquad       i=x,y,z\,, $$ 
 that will be convenient  sometimes for using.   	  
     The curve   ${\boldsymbol{r}}(\tau ,\xi)$ is   reconstructed through  variables $({\boldsymbol q}(0)\,;\, {\boldsymbol p}\,; {\boldsymbol j}(\xi)\,)$  in accordance with the  formula
 \begin{equation}
 \label{z_funct}
  {\boldsymbol{r}}(\tau,\xi)   = {\boldsymbol {q}(0)} + \tau (t_0/ m_0) {\boldsymbol{p}}  +
		           {R_0} \int\limits_{0}^{2\pi}  \left[\,{\xi - \eta}\,\right] {\boldsymbol j}(\tau,\eta) d\eta\,.
   \end{equation} 
	
	As a second step, we must introduce  certain constraints on the set ${\mathcal A}^{\,\prime}$ that define set $\Omega  \subset {\mathcal A}^{\,\prime} $. 
		Criteria for introducing these constraints - the one-to-one correspondence 
	$$ {\mathcal A} \longleftrightarrow  \Omega\,.$$
		It is clear that we must define two constraints, because we introduce three variables  $({\boldsymbol p})$ instead one variable $\Gamma$.

	Fist of all,  we must require that the vectors ${\boldsymbol p}_{\!\perp}$ and ${\boldsymbol f}_{\!\!\perp}$ are proportional.
		In general, it is not true on   set ${\mathcal A}^{\,\prime}$, because these vectors are independent  at this  set.  
		For convenience, let us introduce the complex values: 
	$$ {\sf p} =   (p_{\!\perp})_x + i (p_{\!\perp})_y\,, \qquad     {\sf f} =  ({f}_{\!\perp})_{\,x} + i({f}_{\!\perp})_{\,y}  \,.$$
	In accordance both  the definition of the vector ${\boldsymbol f}_{\!\perp}$ (see (\ref{f_gen})) and  
		the formulas (\ref{j_pm}) we  have the equalities:
	\begin{equation}
        \label{f_perp}
				{\sf f} = -\int_0^{2\pi}\! j_z \sin\xi d\xi  + i \int_0^{2\pi}\! j_z \cos\xi d\xi     =     2\pi {\sf j}_{-1} 				\,.   
				\end{equation}
	Taking into account the replacement  (\ref{rep_factor}),
	formula  (\ref{imp_perp}) takes following form:
	\begin{equation}
	\label{complex_p}
	{\sf p} =    \frac{m_0 \Gamma}{R_0}\, {\sf f}   =     \frac{2\pi m_0 \Gamma}{R_0}\,{\sf j}_{-1}\,.
	\end{equation}

				Therefore,  by virtue of  the formula (\ref{imp_perp}) and the complex-valued notations for the corresponding values, we must demand:
		\begin{equation}
        \label{constr_prop}
			\exists \lambda \in {\sf R}: \qquad \quad  {\sf p} = 2\pi \lambda p_{\,0}  {\sf j}_{-1}\,. 
					\end{equation}
		
	The variables  ${\sf p}$ and ${\sf j}_{-1}$ are the complex numbers here  and  the value      $p_{\,0} = m_0 R_0/t_0 $. 
	As we will show further, the constant  $p_{\,0}$ has a clear physical meaning, so it seems natural to use it as {\it in-put} constant, instead of a value $t_0$.
	However, we will use the constant $t_0$ to simplify some formulas.

	The  condition  (\ref{constr_prop}) can also be written in   following form:	
	\begin{equation}
      \label{constr_0}
	p_x (f_{\!\perp})_y - p_y (f_{\!\perp})_x =0\,. 
	\end{equation}
	Of couse, this equality (just like equality (\ref{constr_prop})) is fullfilled identically   in the case when  our theory is parametrized by the set ${\mathcal A}$.
		In  complex-valued notations the constraint  (\ref{constr_0})  is written as follows:
		
	\begin{equation}
        \label{constr_fin1}
			\Phi_0 \equiv 	{\sf p}\,{\overline{\,\sf j\,}_{-1}} - {\overline{\sf p}}\,{{\sf j}_{-1}}  = 0 \,.
	\end{equation}
	If the quantities  ${\sf p}$   and   ${\sf j}_{-1} $  take non-zero  values and the condition  (\ref{constr_fin1}) is fulfilled, 
			the variable  $\Gamma$  can be determined unambiguously   through the  formulas:
			\begin{equation}
        \label{Gamma}
				\Gamma = \frac{\lambda R_0^2}{t_0}\,, \qquad  \quad    |{\sf p}|^2   - 4\pi^2 p_0^{\,2}\lambda^2 |{\sf j}_{-1}|^2 =0\,.
				\end{equation}
	In corresponding zero points the value  $\Gamma =  \Gamma_0$ is still undetermined.  Let us substitute the representation  (\ref{z_funct}) for the original vortex
	${\boldsymbol{r}_0}(\tau,\xi)$ in LIE equation (\ref{LIE_eq}).  This procedure leads to the equality for the momentum unperturbed  vortex ring:
	$$ p_z = p_{\,0} = m_0 R_0/t_0\,.$$
	To deduce this formula we suppose that $\partial_\tau {\boldsymbol p} = \partial_\tau {\boldsymbol q}(0) =0$.
		These conditions will be coordinated  with the subsequent hamiltonization of our dynamical system.
		In accordance with formula  (\ref{p_z1}) and (\ref{rep_factor}) we have for this case: 		
	 $p_{\,0} = \pi m_0 \Gamma_0 /{R_0}\,.$
	Therefore, $\Gamma_0 = R_0^2/\pi t_0 $.

	Let us return to the perturbed case. In this case the value   $\Gamma$ in the formula (\ref{p_z_pert})       must be same value as was 		
	determined  in formula (\ref{Gamma}).
	That is why we must  	 write the second constraint:
	\begin{equation}
        \label{constr_fin2}
	   \Phi_1 \equiv   |{\sf j}_{-1}|^2 p_z^2  -  (j_\phi^{\,0})^2 |{\sf p}|^2  = 0\,.
		\end{equation}
	
	In special case when  the value  $j_\phi^{\,0} = 0$, we have the  constraint
	\begin{equation}
        \label{constr_fin2_0}
	   \Phi_1 \equiv    p_z   = 0\,
		\end{equation}
	instead constraint  (\ref{constr_fin2}).  This means that set  $\Omega$ describes the planar system here.  
	We will consider the case  $j_\phi^{\,0} = 0 $ only.
		In a general case when $j_\phi^{\,0} \not= 0$,  
	 the set $\Omega  \subset {\mathcal A}^{\,\prime} $  defines by constraints  (\ref{constr_fin1}) and (\ref{constr_fin2}).

	Finally, we have the following
	\begin{prop}
		The variables         ${\boldsymbol j}(\xi)$,  $\boldsymbol{q}$,  $\boldsymbol{p}$, that
                  are declared as the new fundamental variables,  parametrize uniquely considered dynamical system. These variables are constrained by the equalities  (\ref{constr_fin1})   and (\ref{constr_fin2}).
		\end{prop}

	\section{Energy and Hamiltonian structure }

						The straightforward calculation   of the energy of a vortex filament is usually performed using the  canonical formula \cite{Saffm}
       \begin{equation}
  \label{can_energy}
       {\mathcal E} = \frac{1}{8\pi}\,\iint
       \frac{\boldsymbol{\omega}(\boldsymbol{r})\boldsymbol{\omega}(\boldsymbol{r}^{\prime})}{|\,\boldsymbol{r} - \boldsymbol{r}^{\prime}|}\,dVdV^{\prime}=
       \frac{{\Gamma}^{\,2}}{8\pi}\iint 
       \frac{\partial_\xi{\boldsymbol{r}}(\xi)\partial_\xi{\boldsymbol{r}}(\xi^{\prime})}{|\,{\boldsymbol{r}}(\xi) -  {\boldsymbol{r}}(\xi^{\prime}) |}\,d\xi d\xi^{\prime}\,,\nonumber
       \end{equation}
			The result is   unsatisfactory if  the filament has  zero thickness:
			the integral in this formula   diverges.  
	The standard approach to solve this problem  is to take into account the finite thickness $a$ of the filament
						and  the subsequent  regularization  of  the integral  (see, for example, \cite{Zhu}, where the interaction between pairs of quantized vortex rings was studied).

	In the proposed approach, we have chosen a different method: the energy of the arbitrary   configuration  in our model will be   considered     from the group-theoretical viewpoint.
	Indeed,           the Lee algebra of group $\widetilde{\mathcal G}_3$ has three Cazimir functions:  
	
	 $$ {\hat C}_1 = m_0 {\hat I}\,,\quad 
  {\hat C}_2 = \left({\hat M}_i  - \sum_{k,j=x,y,z}\epsilon_{ijk}{\hat P}_j {\hat B}_k\right)^2 
  \quad {\hat C}_3 = \hat H -  \frac{1}{2m_0}\sum_{i=x,y,z}{\hat P}_i^{\,2}\,,$$                         
       where        ${\hat I}$ is the unit operator,     ${\hat M}_i$,   $\hat H$,  ${\hat P}_i$         and  ${\hat B}_i$  ($i = x,y,z$)
        are the respective generators of rotations, time and space translations and Galilean boosts. 
		 As it is well known, the function  ${\hat C}_3 $  can be interpreted as  an  ''internal energy of the particle''. 
	Because our dynamical system has an ''internal degrees of the freedom'',  the function  ${\hat C}_3 $ can depend on the internal variables.
		We define these functions as follows

       $${ C}_3  			=   {\mathcal E}_0 \sum_{n>1} |\,{\sf j}_{\,-n}|^2 n\sqrt{n^2 -1}			\,.$$         			
				Here we have introduced into consideration the 						
						value ${\mathcal E}_0= m_0 R_0^2/t_0^2$ which defines the energy scale   in our theory. 
						The choice of the function $C_3$ will be quite justified after the definition of the Hamiltonian structure.

					As a result, the following function on the set $\Omega$ is a good candidate
					for the energy\footnote{We are considering here  the energy of excitations only.}:                
 \begin{equation}
  \label{energy_1}
 {H}_0(p_1,p_2,p_3\,;{\sf j}) = \frac{{\boldsymbol{p}}^{\,2}}{2m_0}   +  
 {\mathcal E}_0 \sum_{n>1} |\,{\sf j}_{\,-n}|^2 n\sqrt{n^2 -1}	\,.
\end{equation}                
  To complete the consideration of energy, we must define the Poisson brackets that are  compatible with the dynamics and constraints.  		
		
		In this article we consider the simplest case that corresponds the value  $j_\phi^{\,0} = 0$. Consequently, the constraint  (\ref{constr_fin2_0}) is fulfilled.
		It is quite natural to add the additional constraint
		\begin{equation}
  \label{add_constr}
	  \Phi_2   \equiv     q_z   - R_0 \tau   =   0 \,.
	\end{equation}

		Pursuant to Dirac's  prescriptions about the primacy of  Hamiltonian structure,
  we define such structure  axiomatically here.  The correspondent definitions are following.
 
  \begin{itemize}
  \item Phase space ${\mathcal H} =  {\mathcal H}_3  \times  {\mathcal H}_j   $. The space $ {\mathcal H}_3$ is the phase space of a $3D$  free structureless    particle.
	It is  parametrized by the variables 
   ${\boldsymbol{q}}$ and  ${\boldsymbol{p}}$.  The space    $ {\mathcal H}_j$ is parametrized by the quantities
	 	${\sf j}_{\,-n}$ ($n  = 0, 1, \dots$).		
		 \item Poisson structure:
  \begin{eqnarray}
  \{p_i\,,q_j\} & = & \delta_{ij}\,,\qquad i,j = x,y,z\,, \nonumber \\
  \label{ja_jb}
  \{ {\sf j}_{\,m}, \overline{\,\sf j\,}_{\,n}\} & = & (i/{\mathcal E}_0 t_0)\, \delta_{mn}\,, \qquad m,n = -1,-2,\dots
  \end{eqnarray}
  
	All other brackets  vanish. 		The variable  ${\sf j}_{\,0}$ 	 annulates all brackets. Thus, the Poisson structure of the theory is degenerate in general: the value ${\sf j}_{\,0}$  marks the symplectic sheets  where the structure will be non-degenerate. 
					
		\item  Constraints   (\ref{constr_fin1}),  (\ref{constr_fin2_0})  and (\ref  {add_constr}).  It is clear that constraints  (\ref{constr_fin2_0})  and (\ref  {add_constr}) form the pair 
		of second type constraints in Dirac terminology.  Moreover the following  equalities  hold:
		$$ \{\Phi_0, \Phi_k\} = 0\, \qquad k = 1,2.$$
		Therefore, we can exclude the coordinates $p_z$ and $q_z$ from the phase space $ {\mathcal H}_3$ replacing it to the phase space $ {\mathcal H}_2$. The last one - the phase space of a   free structureless    particle on a plane.  There are no additional constraints here because  the equality
			$ \{H, \Phi_0\} =   0\,.$
  \item Hamiltonian 
	\begin{equation}
	\label{H_ful}
	H=H_0+  \ell\Phi_0\,,
	\end{equation}
	where  the function  $H_0  $ is defined by the formula (\ref{energy_1}) with replacing   $\boldsymbol{p}  \to  \boldsymbol{p}_\perp$.
    The  quantity  ${\ell}$ is the Lagrange factor.
  \end{itemize}

					Let's pay attention to the following point.  We introduce the set of the new variables ${\mathcal A}^{\,\prime}$ which is more extensive than the set of original variables. Constraints on set 
					 ${\mathcal A}^{\,\prime}$ were postulated.  These constraints lead to a certain arbitrariness in dynamics.  That is why the constructed  dynamical system is not equivalent to the original one, but it is more extensive.  Indeed, 		
					let us define the physical (dimensional) time $t = t_0\tau$. The following Proposition is true:
		\begin{prop}
		The following Hamilton equations are valid:
	\begin{eqnarray}
				\frac{\partial \boldsymbol{q}}{\partial t }  & = & \{H_0,{\boldsymbol{q}}\} = \frac{\boldsymbol{p}}{m_0} \,, \qquad 
							\frac{\partial \boldsymbol{p}}{\partial t}  = \{H_0, \boldsymbol{p}\}  = 0\,,\\
								\frac{\partial {\sf j}(\tau,\xi) }{\partial t}	& = & \{H_0, {\sf j}(\tau,\xi)\}  =  
								\frac{i}{t_0} \sum_{|n| >1} {\sf j}_{\,n}n\sqrt{n^2 - 1}\, e^{\,i[\,n\xi  +  n\sqrt{n^2 - 1}\,\tau\,]} \,.
				\end{eqnarray}	
			\end{prop}		
					Because the Hamiltonian $H$ differs from $H_0$, constructed system is equivalent to original if the Lagrange factor  ${\ell} = 0$.

				\section{Quantization } 
				
				Numerous articles devoted to the problem of turbulence show that the understanding of this phenomenon is still not full. This statement also fully applies to the turbulence of quantum liquids. Without setting out to review the literature on this issue, we note that a number of authors (see, for example, \cite{TsFuYu} ) assume that the key to understanding this problem is the topological defects of such liquids - vortices. That is why the quantum description of vortices is an important task. The information about the spectrum of energy of such defects for concrete tasks allows investigating certain statistical characteristics of the system.  In this paper, we aim to develop a new approach to the quantization of a single closed vortex with zero thickness. The author believes that in the future, the results obtained may provide new opportunities for explaining the behavior of quantum liquids.

					The constructed Hamiltonian structure  opens up possibilities for quantization of the small perturbations of the vortex ring under study.
										Firstly, we must define a Hilbert space $\boldsymbol{H}$   of the quantum states of our dynamical system.  The structure of the phase space ${\mathcal H}$ lead to the 
					following structure:
					\begin{equation}
	\label{space_quant}
	\boldsymbol{H}  =  \boldsymbol{H}_2 \otimes   \boldsymbol{H}_F\,,
	\end{equation}
			where the symbol   $\boldsymbol{H}_2$  denotes the Hilbert space  of a free structureless particle on a plane (the space 			
			$L^2({\sf R}_2)$ for example) and  symbol $\boldsymbol{H}_F $ denotes the Fock space for the infinite number of the harmonic oscillators.		 
				The creation and annihilation operators which are defined in the space 	$\boldsymbol{H}_F $, have standard commutation relations
				$$ [\,\hat{a}_m, \hat{a}_n^+] = \hat{I}_F\,, \qquad \hat{a}_m|\,0\rangle = 0  \,, \qquad  m,n = 1,2,\dots \,, \qquad    |\,0\rangle \in    \boldsymbol{H}_F           \,,$$   
				where the operator   $\hat{I}_F$  is unit operator in the space    $\boldsymbol{H}_F $.
								Let us quantize our theory.  We must to construct the function $A \to \hat{A}$, where  $A$ denotes some classical variable and $\hat{A}$ denote some operator in the space $\boldsymbol{H}$.
					Traditionally,  we must to demand
					$$  [\hat{A},\hat{B}]  = -i\hbar \widehat{\{A,B\}}\,$$
					if  the quantities $A$, $B$, $\dots$ denote  the funamental variables in our theory.
					This equality can possess of some  ''anomalous terms''  if the ''observables'' $A$, $B$ are the functions of the fundamental variables.
					These terms depend  on a rule of the ordering of  non-commuting operators.   We will not discuss these issues here \cite{Berezin}.

					Let us consider the case when $\boldsymbol{H}_2  = L^2({\sf R}_2)$.  This case corresponds to the perturbation of a vortex ring in unbounded space.
					Our postulate of quantization is following:
					$$ q_{x,y} \to  q_{x,y}\otimes \,\hat{I}_F  \,,\qquad  p_{x,y} \to - i\hbar\frac{\partial}{\partial q_{x,y}} \otimes \,\hat{I}_F \,,\qquad
					{\sf j}_{\,-n} \to \sqrt{\frac{\hbar}{t_0{\mathcal E}_0}}\, (\hat{I}_2 \otimes\,\hat{a}_n)\,, $$
					where  $ n = 1,2,\dots\, $ and  operator $\hat{I}_2$  is unit operator in the space  $\boldsymbol{H}_2$.
Next, we will not write the index $n = 1$ explicitly: $\hat{a}_{1}   =  \hat{a}$ and so on. This simplification will be justified later.
Moreover  we will not write the constructions $(\dots  \otimes \,\hat{I}_F)$  and  $ (\hat{I}_2 \otimes\,\dots)$ explicitly, hoping that this will not lead to misunderstandings.

As a next step, we should to construct the physical subspace  $\boldsymbol{H}_{phys} \subset    \boldsymbol{H}$. 
 In accordance with Dirac's prescription,  the presence  of constraint (\ref{constr_fin1}) leads to  the equation for the vectors $ |\psi\rangle  \in \boldsymbol{H}_{phys} $:

$$   {\widehat\Phi_0}|\psi\rangle  = 0 \,.$$

Here it is extremely important to pay attention to the following fact.
					In a classical theory, all forms of the first-type constraints lead to the same theory: in our case, we can assume $\Phi_0 = 0$ or $\Phi_0^{\,2} = 0$ and so on.
					This is not the case at all in the quantum version of the model.
					The different forms of first type constraints 
					correspond  to the different equations for the  ''physical vectors''  in a quantum theory\footnote{Apparently, Dirac's words are still relevant: 
					{\it''\dots methods of quantization are all of the nature of practical rules, whose application depends on consideration of simplicity''}}. 
					For instant,  it is  clear that the solutions of the equation ${\widehat\Phi_0}|\psi\rangle  = 0$  differ from the solutions of the equation 
					${\widehat\Phi_0^{\,2}}|\psi\rangle  = 0$.  
Consequently, we   need to supplement the  quantization rules with a specific choice of the form of the constraint in  classical theory.

Let us investigate this problem in our model in more detail.  First of all, we consider the classical constraint in form  (\ref{constr_prop}). 
Taking into account formula    (\ref{f_perp}),  
 we search the vectors $|\psi_{phys}\rangle$ so that

\begin{equation}
        \label{constr_q_pr1}
			\exists \lambda \in {\sf R}: \qquad  (\hat{\sf p} - 
			2\pi\lambda  p_{\,0}\sqrt{\frac{\hbar}{t_0{\mathcal E}_0}}\,\hat{a}) |\psi_{phys}\rangle  =0  \,.	
					\end{equation}

Let   the complex number  ${\sf p} = p_x + i p_y$ is  eigenvalue  that corresponds to the (generalized) eigenvector   
  $|{\sf p}\rangle \in \boldsymbol{H}_2^{\,\prime}$  of  the operator $\hat{{\sf p}}$. The notation $\boldsymbol{H}_2^{\,\prime}$ means  that   prosedure of a rigging of the space $\boldsymbol{H}_2$ must be fulfilled  	to consider the generalized eigenvectors rigorously \cite{BerShu}.  
As ansatz for the solutions of the equation (\ref{constr_q_pr1}), we use  the following  form  for the ''physical   vectors''       $|\psi_{phys}\rangle$:  
\begin{equation}
\label{phys_vect1}
|\psi_{phys}({\sf p})\rangle  =  |{\sf p}\rangle |\psi_p\rangle\,,\qquad |{\sf p}\rangle\in \boldsymbol{H}_2^{\,\prime}\,,\quad |\psi_p\rangle\in\boldsymbol{H}_F\,.
\end{equation}

Therefore, the equation for the vector  $|\psi_p\rangle$ takes following form:

\begin{equation}
        \label{constr_q_pr2}
			\exists \lambda \in {\sf R}: \qquad  ( {\sf p} - 
			2\pi\lambda p_{\,0}\sqrt{\frac{\hbar}{t_0{\mathcal E}_0}}\,\hat{a}) |\psi_{p}\rangle  =0  \,, 
			\quad |\psi_p\rangle\in\boldsymbol{H}_F\,,  \quad  {\sf p}  \in {\sf C}  \, .	
					\end{equation}

In other words, the  vectors $|\psi_{p}\rangle$  are the eigenvectors of the  spectral problem   (\ref{constr_q_pr2}).

\begin{prop}
Let the the vectors $| {\sf p}/\lambda  \rangle \in  \boldsymbol{H}_F$ form the system of coherent states\footnote{See, for example,  \cite{Perelomov}} for  Heisenberg - Weyl group with algebra
$$ [\,\hat{a}, \hat{a}^+] = \hat{I}\,, \qquad [\,\hat{a}^+, \hat{I}] = [\,\hat{a}, \hat{I}] = 0\,$$
as follows:
\begin{equation}
\label{coherent_1}
| {\sf p} /\lambda \rangle = 
\exp\Bigl[   ({\sf p}\,\hat{a}^+    - {\overline{{\sf p}}}\,\hat{a}) /
2\pi\lambda \sqrt{\frac{\hbar p_{\,0}}{R_{\,0}}}\, \Bigr]\,|0\rangle\,, \qquad \lambda \in {\sf R}\,, \quad {\sf p} \in {\sf C}\,.
\end{equation}
Then the equation (\ref{constr_q_pr2}) has solutions
\begin{equation}
\label{eig_ham}
 |\psi^\lambda\rangle =  |n_1,\dots,n_k; {\sf p}/\lambda\rangle  \equiv
   \hat{a}^+_{n_1}\hat{a}^+_{n_2}\dots\hat{a}^+_{n_k}| {\sf p}/\lambda  \rangle \,, \quad  \qquad  n_j > 1 \,.\nonumber
\end{equation}
\end{prop}
This statement can be verified directly because the  coherent states 
 $| {\sf p}/ \lambda \rangle$  are eigenvectors of the operator  $\hat{a}$
with eigenvalue ${\sf p}/(2\pi\lambda \sqrt{{\hbar p_{\,0}}/{R_{\,0}}}) $.
  Recall that the parameter $\lambda$ has the meaning of ''dimensionless circulation'' in our model.

Let us consider the vectors  $|\psi_{[n]}^\lambda({\sf p})\rangle \in \boldsymbol{H}^{\,\prime}$:

\begin{equation}
\label{ent_2}
|\psi_{[n]}^\lambda({\sf p})\rangle  =   C_{[n]}   
|{\sf p}\rangle |n_1,\dots,n_k; {\sf p}/\lambda\rangle \,, \qquad  \lambda  \in  {\sf R}  \,,\quad [n] =   n_1,\dots,n_k  \,,
\end{equation}
 where numbers  $C_{[n]} $  are normalizing coeffisients, so that normalizing conditions are fulfilled:
$$   \langle  \psi_{[n]}^\lambda({\sf p})  |\psi_{[m]}^\lambda({\sf p^{\,\prime}})\rangle  = \delta_{k l} \delta_{n_1 m_1}\dots \delta_{n_k m_k} 
 \delta(p_x - p_x^{\,\prime})\delta(p_y - p_y^{\,\prime})\,.$$

Here we should note that  the entire set of vectors (\ref{ent_2})
can't form a physical subspace $\boldsymbol{H}_{phys} \subset    \boldsymbol{H}$ because 
any superpositions
$$  c_1 |\psi_{[n]}^{\lambda_1}({\sf p}_1)\rangle  +  c_2 |\psi_{[n]}^{\lambda_2}({\sf p}_2)\rangle \,,\qquad  \lambda_1 \not= \lambda_2\,,\quad  {\sf p}_1 = {\sf p}_2  $$
 is not a solution to the spectral problem (\ref{constr_q_pr2}).  Moreover we can't postulate ''superselection  rules'' here.
Indeed, the  coherent states $| {\sf p}/ \lambda \rangle$  are not orthogonal for different values of parameter ${\sf p}/\lambda$. Therefore, 
the vectors $|\psi_{[n]}^\lambda({\sf p})\rangle$  are not orthogonal for same values  ${\sf p}$  and different values    $\lambda$.

We will proceed as follows. 
Let the vector $|\psi_{[n]}^\star({\sf p})\rangle  \in \boldsymbol{H}^{\,\prime} $ will be the vector  (\ref{ent_2}) where 
the value  $\lambda = \lambda_{\,0} =1/\pi$:

\begin{equation}
\label{ent_star}
|\psi_{[n]}^\star ({\sf p})\rangle  =  C_{[n]}    
|{\sf p}\rangle |n_1,\dots,n_k; {\sf p}/\lambda_{\,0}\rangle\,,  \qquad k = 0,1,2,\dots         \,.
\end{equation}

The state (\ref{ent_star}) which corresponds $k = 0$, we call as ''ground state'' of our theory. 
Recall that the  value   $\lambda  = \lambda_{\,0} $  corresponds to the  circulation $\Gamma_0$  of the unperturbed vortex ring: $\Gamma_0 = {R_0^{\,2}}/{\pi t_0 }$.

We declare that the physical subspace  $\boldsymbol{H}_{phys}$ is spanned by the following vectors
\begin{equation}
\label{phys_vect}
|\psi_{phys}\rangle =    \sum_{n_1,\dots,n_k \atop n_j>1}  \int  dp_x d p_y\,\varphi_{n_1,\dots,n_k} (p_x,p_y) 
|\psi_{[n]}^\star ({\sf p})\rangle \,,
\end{equation}
where the wave functions  $\varphi_{n_1,\dots,n_k} (p_x,p_y)$    are normalized.

Does the constructed space  $\boldsymbol{H}_{phys}$   describes the states with circulation $\Gamma_0$ only?  In our opinion, 
the properties of coherent states allow us to assume that this is not the case.
Indeed, every  system of coherent states   $|\alpha\rangle$ is an overdetermined system. Because $\langle\alpha_1  |\alpha_2\rangle \not= 0$ even for different complex numbers
$\alpha_1$  and $\alpha_2$, we can conclude that any  specific coherent state $|\alpha_0\rangle$  contains ''some part'' all other coherent states $|\alpha\rangle$,  $ \alpha \not= \alpha_0$.

Returning to our notation, we have the following expression for the amplitude  for any $\lambda \in {\sf R}$:
 \begin{equation}
\label{amplitude}
\langle  \psi_{[n]}^\lambda({\sf p})  |\psi_{phys}\rangle = 
\varphi_{[n]} (p_x,p_y)\exp\left[- \frac{|{\sf p}|^2 R_{\,0}}{8\, \hbar\, p_{\,0}}\left(\frac{\lambda_{\,0}}{\lambda} - 1\right)^{\!2}  \right]\,.
\end{equation}
Although  $|\psi_{[n]}^\lambda({\sf p})\rangle \not\in \boldsymbol{H}_{phys} $  for the case $\lambda \not= 1/\pi$, we suppose that the  amplitude (\ref{amplitude})
defines the probability density to find our dynamical system with circulation   $\Gamma = {\lambda R_0^2}/{t_0}$, transverse impulse $ {\sf p} = p_x + i p_y$ and quantum numbers
$[n] = \{n_1,\dots,n_k\}$.

The numbers   $[n] $ define the energy of our system.  Indeed, 
in accordance with our classical formulas for the  Hamilton function, the quantum expression   for the  Hamiltonian  $\hat{H}$  takes the form 
\begin{equation}
\label{H_quant}
\hat{H} = \frac{\hat{{\sf p}}^+\hat{{\sf p}}}{2m_0}  +  \frac{\hbar}{t_0} \sum_{n>1} \hat{a}^+_{n} \hat{a}_{n}  n\sqrt{n^2 - 1}	\,.\nonumber
\end{equation}
 The following statement can be proved by direct verification:

\begin{prop}
The vectors $|\psi_{[n]}^\star ({\sf p})\rangle  =  C_{[n]} |{\sf p}\rangle|n_1,\dots,n_k; {\sf p}/\lambda_{\,0}\rangle  $    are the eigenvectors of operator  $\hat{H}$ with eigenvalues
\begin{equation}
\label{spectr_ham}
{\mathcal E}  =   \frac{|{\sf p}|^2}{2m_0} +  \frac{\hbar}{t_0} \sum_{n>1} \sum_{j=1}^k \delta_{n,n_j} n\sqrt{n^2 - 1} \,.
\end{equation}
\end{prop}

As regards the constraint in the form (\ref{constr_fin1}),
the following equality takes place:
$$ \langle \psi_{phys} | {\widehat\Phi_0} |\psi_{phys}\rangle  = 0\,.$$ 

As a result, the constructed quantum description of the system allows for a visual interpretation of the vortex as a structured particle. 
The connection between external degrees of freedom (space $\boldsymbol{H}_2$) and internal degrees of freedom  (space $\boldsymbol{H}_F$)
 is nontrivial due to the constraint   (\ref{constr_q_pr1}). As a consequence, the quantum states of such a system are entangled states.
The formula (\ref{phys_vect}) demonstrates that  the vectors

\begin{equation}
|\psi_{[0]}^\star ({\sf p})\rangle  =      
|{\sf p}\rangle |{\sf p}/\lambda_{\,0}\rangle\,  \nonumber     
\end{equation}
forms the ''set of the ground states'' of our system.
  These vectors 
corresponds to the vortex ring with the exact value of the transverse momentum ${\sf p}$,  the most probable  value of the circulation $\Gamma_0 = {R_0^{\,2}}/{\pi t_0 }$
and the minimal energy  ${|{\sf p}|^2}/{2m_0}$.

\section{Concluding remarks }

This paper has constructed a model describing the classical and quantum dynamics of small perturbations of the vortex ring, which evolves according to the LIE equation.
The theory has three-dimensional constants: $R_0$ (radius of unperturbed vortex ring), $p_0$ (momentum of unperturbed vortex ring), and $m_0$ (the central charge for the central extension of Galilei group).  We quantized our model as an abstract dynamical system,   without any connection with the quantum properties of the surrounded liquid.  Of course, taking into account such property is an important area that is presented in the literature (see \cite{Stamp}, for example).
In the case under consideration, when $\boldsymbol{H}_2  = L^2({\sf R}_2)$, 
 the eigenvalues of the momentum operator belong to a continuous set.  The same can be said about the energy ${\mathcal E}$ (see (\ref{spectr_ham})) and the
 circulation $\Gamma$ (see  (\ref{amplitude})).
However, there is no contradiction with the experiment: for example, all the experiments, where the quantization of the circulation was observed  \cite{Donn}, correspond to one of the real cases when 
certain boundary conditions are present.
To take into account any boundary conditions in our approach,  we should consider the  case  when $\boldsymbol{H}_2  = L^2({\sf D})$, where the domain  ${\sf D}$ is a certain compact subset
 of a plane ${\sf R}_2$.
We suppose that corresponding theory leads to the discrete values of energy ${\mathcal E}$ and circulation $\Gamma$.  The author hopes to devote the next article to this issue.

\end{document}